\documentclass[a4paper,10pt,twoside]{cpc-hepnp}

\usepackage{multicol}
\usepackage{graphicx}
\usepackage{booktabs}
\usepackage{amssymb,bm,mathrsfs,bbm,amscd}
\usepackage[tbtags]{amsmath}
\usepackage{lastpage}
\usepackage{CJK}

\begin{document}
\begin{CJK*}{GBK}{song}

\fancyhead[c]{\small Chinese Physics C~~~Vol.  , No. 1 (2022)
000001} \fancyfoot[C]{\small 010201-\thepage}

\footnotetext[0]{Received   2022}

\title{Seniority and $(\frac{9}{2})^n$ configurations in neutron-rich Nickel isotopes \thanks{This research has been supported by the Interdisciplinary Scientific and Educational School of Moscow University "Fundamental and Applied Space Research"}}

\author{
    S. Sidorov$^{1,2}$,%
\quad D. Zhulyaeva$^{1}$,
\quad T. Tretyakova$^{1,2}$\email{tretyakova@sinp.msu.ru}
}
\maketitle

\address{%
{$^1$
Faculty of Physics, Lomonosov Moscow State University, Moscow 119991, Russia}\\
{$^2$Skobeltsyn Institute of Nuclear Physics, Lomonosov
Moscow State University, Moscow 119991, Russia}\\
}

\begin{abstract}
Excited states in low-energy spectra in $^{70-76}$Ni are considered. To this end, pairing forces in form of surface delta interaction are employed to account for formation of the ground state multiplet with seniority $\nu = 2$ states. The multiplet splitting is described with mass relations of masses of neighbouring nuclei. Subsequently, seniority model is used to reproduce or predict the states $\nu = 3$ in odd-even isotopes and $\nu = 4$ in even-even isotopes. Correct account of $2_1^+$ state should allow for description of reversed order of states $J = 4$ with $\nu = 2$ and $\nu = 4$ observed in experiment. The results obtained are compared with the structure of similar multiplets in $N=50$ isotones.

\end{abstract}

\begin{keyword}
shell model, nucleon pairing, seniority scheme, neutron-rich isotopes
\end{keyword}

\begin{pacs}
21.10.-k, 21.60.Cs, 27.50.+e
\end{pacs}

\footnotetext[0]{\hspace*{-3mm}\raisebox{0.3ex}{$\scriptstyle\copyright$}2022
Chinese Physical Society and the Institute of High Energy Physics
of the Chinese Academy of Sciences and the Institute
of Modern Physics of the Chinese Academy of Sciences and IOP Publishing Ltd}%

\begin{multicols}{2}

\section{Introduction}

Last decades have seen nickel isotopes, especially those with excess of neutrons, receive special attention due to various applications in both theoretical and experimental nuclear physics \cite{CCC17, Wat19}. The chain of $N=40$ neutron-rich isotones and $^{68}$Ni in particular, as well as other neighbouring isotopes, present another case of \textit{an island of inversion} --- a region with significant changes in shell structure and abrupt transitions from spherical to deformed states of atomic nuclei, similar to an island of inversion in $N=20$ region. Doubly magic $^{78}$Ni is of particular interest when it comes to studies of magic number conservation and shape co-existence. This is the only neutron-rich magic nuclide on the very edge of nuclear chart that is currently available in experiment \cite{TSD19}.

Nuclei in vicinity of $^{78}$Ni are of great significance in astrophysics \cite{GLM07}, serving as some of the initial isotopes produced in the r-process responsible for production of nuclei with $Z>26$. The probability of synthesis of different chemical elements along different neutron-rich isotope chains strongly depends on the shell structure of these isotopes.

The spectroscopic studies of $^{78}$Ni and its neighbours are included in the program of all the next-generation RI-beam in-flight facilities such as Radioactive Isotope Beam Factory (RIBF) in Japan, Facility for Rare Isotope Beams(FRIB) in USA, Facility for Antiproton and Ion Research (FAIR) in Germany. In just the past two decades, major progress was made in experimental studies in isotopes $^{70-76}$Ni \cite{GGL97, GBB98, SGM03, CWS11, M16, MBW18, GGM20, GAD20, EMN21}. One particular aspect broadly discussed is the conservation seniority number $\nu$ in these and other nuclei with $j=9/2$ filled shell \cite{RR01, RR03}. Earlier, a series of experiments (e.g. \cite{BFK95, MBF99}) was designed to verify the magicity of the $N=40$ shell closure in $^{68}$Ni and the isolation of $\nu g_{9/2}$. As a result, a hypothesis was made regarding the existence of excited states that should typically form as a consequence of pairing. These states should be isomeric to those well known in $N=50$ isotones \cite{GBB98}. Detailed analysis showed, however, that magicity of the $N=40$ gap is less defined compared to $Z=40$ gap \cite{MBF99}. Further accumulation of experimental data on nickel isotopes with greater neutron excess had us acknowledge the differences between formation of low-lying excited states in $\nu = 4$ nuclei $^{72}$Ni, $^{74}$Ni and $^{94}$Ru, $^{96}$Pd, as well as raise the question regarding the order of levels with the same value of $J^\pi$ and different seniority \cite{Isacker}. New spectroscopic data on $^{72}$Ni, $^{74}$Ni let us examine this matter more thoroughly \cite{M16, MBW18}.

We consider the spectra of excited states in neutron-rich $^{70-76}$Ni isotopes and $N=50$ isotones in a simplistic phenomenological approach. The seniority $\nu=2$ part of the spectrum is described using the $\delta$-force approximation for interaction of two like nucleons on the $j=9/2$ shell. The only parameter in this case, which is the ground state multiplet (GSM) splitting, can be estimated using the mass relations for pairing energy. States with higher seniority number can be then reproduced within the frame of seniority scheme. The advantages of such an approach lie with its simplicity and lack of free parameters. Earlier it was used to calculate the GSM in nuclei near $^{208}$Pb, where reasonable description of high momentum states was achieved \cite{Step,SI18}.

\section{Ground state multiplets and seniority scheme}

Nucleon pairing brings about the formation of specific sets of excited states that constitute the ground state multiplet. In case of a pair of like nucleons in $j$ state, the spins of these states take on even values and are characterized by the isospin of the nucleon pair $T=1$ and spin $S=0$. In seniority model, these excited states are degenerate \cite{R42, Bayman}. In 1950, Goeppert Mayer proposed the description of nucleon pairing $\delta$-potential approximation \cite{GM50}. In this case, the relative shift of the excited state $J$ can be found as \cite{DS53}:
\begin{align}
	\frac{\Delta E_{J}}{\Delta E_{0}}=(2j+1)
	\begin{pmatrix}
		j& j& J\\
		\frac{1}{2}& -\frac{1}{2}& 0
	\end{pmatrix}^2.
\end{align}

If we know the pairing energy $\Delta E_0=\Delta_{NN}$, the energy of all GSM states $E_J$ can be calculated as:
\begin{align}
	E_J = \Delta_{NN}(1 - \frac{\Delta E_{J}}{\Delta E_{0}}).
\end{align}

A picture of nucleon pairing depicted above is based on a description of a typical nucleus with two nucleons above the doubly magic core. On the example of two neutrons above the core $(N-2, Z)$, the lack of pairing would result in equality between two neutron separation energy $S_{nn}(N, Z)$ in nucleus $(N, Z)$ and double the separation energy $S_n(N-1, Z)$. Pairing correlations then result in difference between these two values, and the corresponding energy of residual interaction can be found using the mass relation \cite{Pr62}:
\begin{align}
	\nonumber &\Delta_{nn}(N,Z) = (-1)^{N} [S_{nn}(N,Z) - 2 \cdot S_n(N-1,Z)]\\
	\nonumber &= (-1)^{N} [B(N,Z) - 2 \cdot B(N-1,Z) + B(N-2,Z)]\\ 
	&= (-1)^{N} [S_n(N,Z) - S_n(N-1,Z)].
	\label{delta}
\end{align} 

Here, $B(N,Z)$ is the binding energy of (N,Z). In case of protons, the $n$ indexes are swapped for $p$ and the relation is based on masses or binding energies of isotones. Such an estimate requires the data on three neighbouring nuclei. The question of correspondence between pairing energy and different mass relations, related to even-odd staggering, was investigated thoroughly, e.g. in \cite{BM71,JHJ84,MN88,Do01}). Earlier we conducted the analysis of different mass relations based on masses of three or more isotopes \cite{CPC17}. Splitting of GSM for two nucleons above the closed core and the pairing energy are strongly correlated.  

	\begin{center}
		\includegraphics[width=90mm]{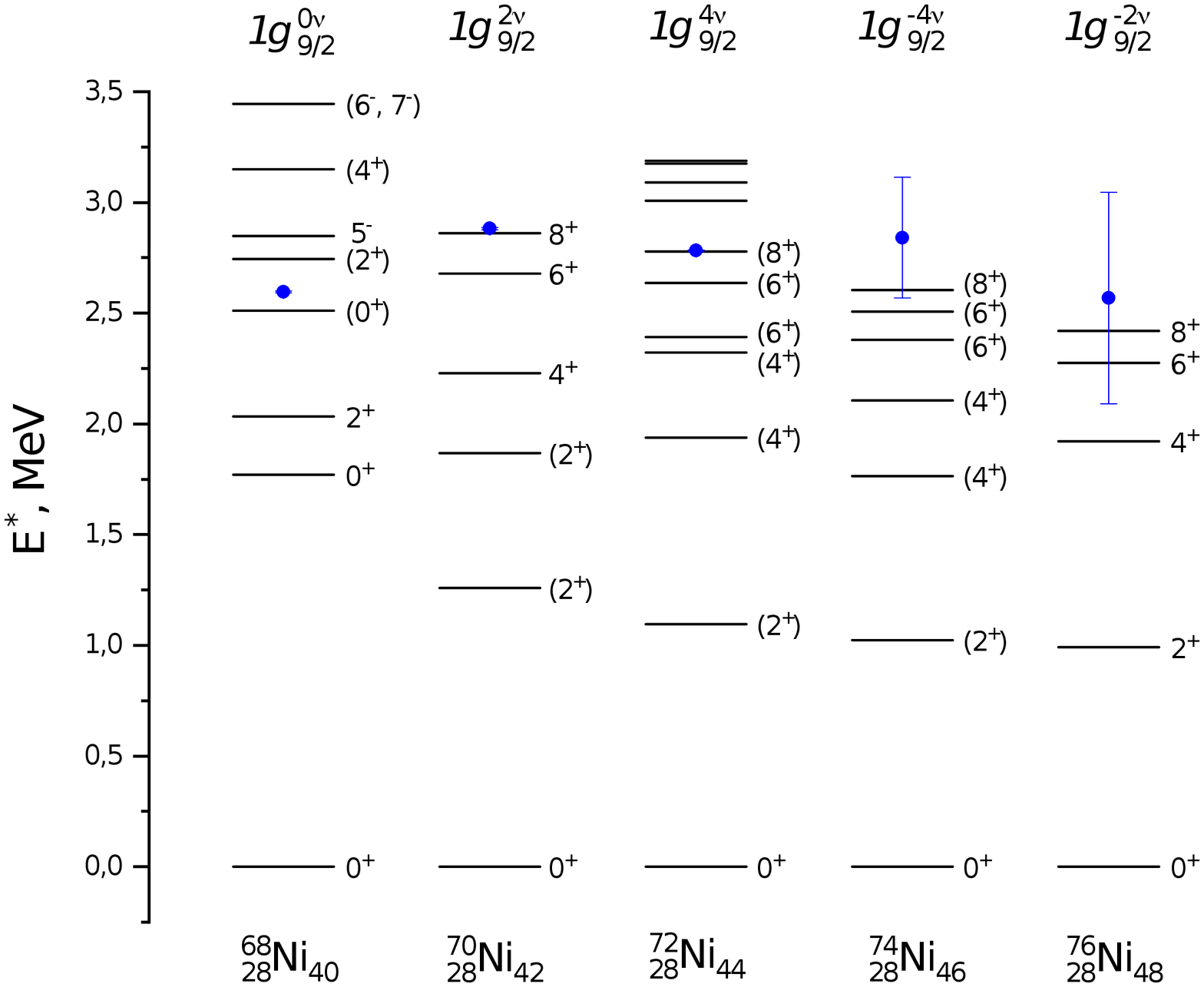}   
		\includegraphics[width=90mm]{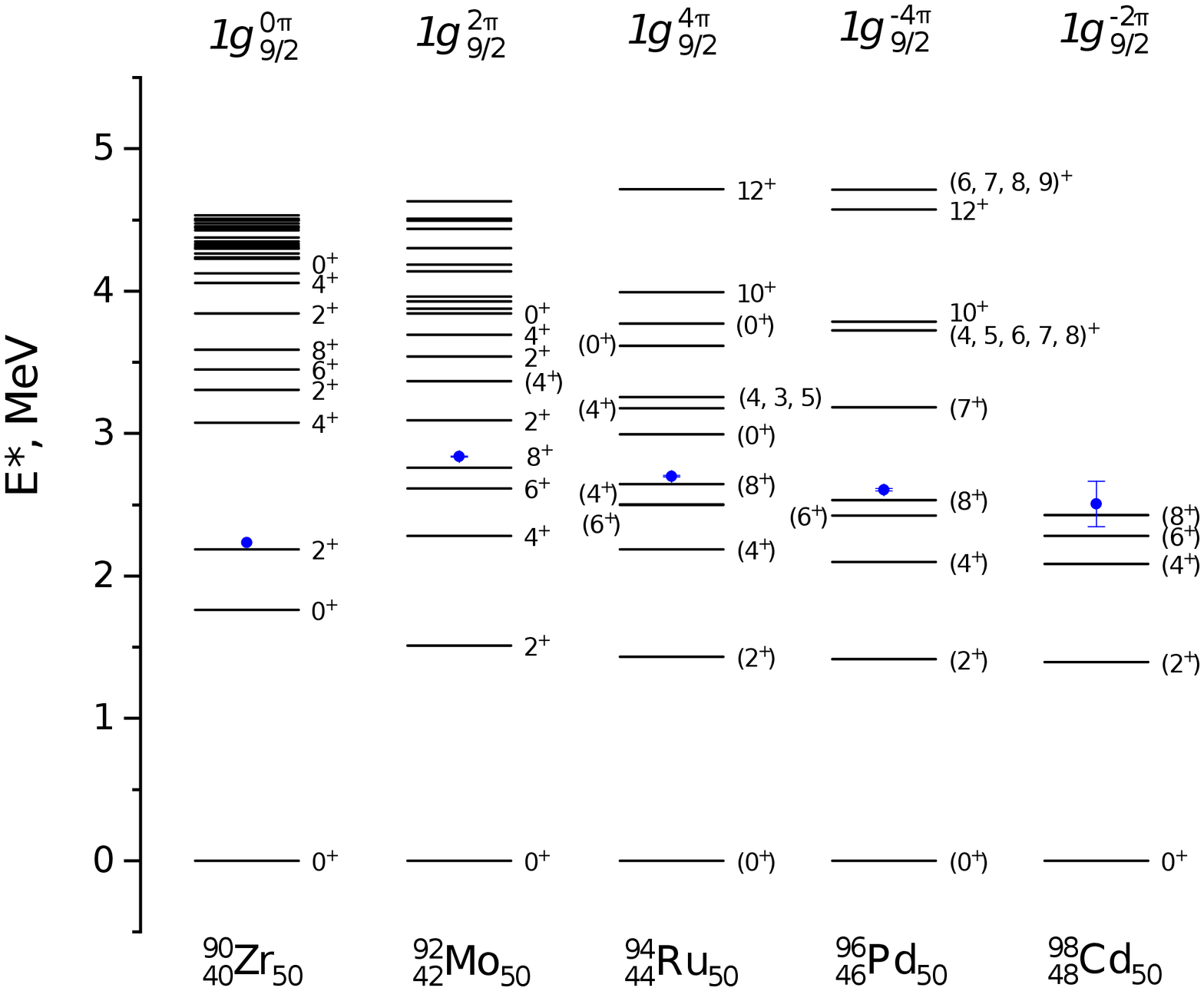} 
				\figcaption{Experimental spectra: a) even isotopes $^{68-76}$Ni and b) even isotones $N=50$ with filled $j=9/2$ subshell. The dots represent the value of neutron pairing energy $\Delta_{nn}^{(4)}$ in nickel isotopes or proton pairing energy $\Delta_{pp}^{(4)}$ in isotones $N=50$.}
		\label{exp}
	\end{center}
In \cite{PEPAN17} it was shown, that for nuclei near the line of stability the best fit between these values obtains for mass relations from \cite{MN88}, based on five neighbouring isotopes.  Nickel isotopes under cosideration in present paper are located close the neutron drip-line, so the usage of additional data is undesirable, so calculations of GSM in neutron-rich nickel isotopes and $N=50$ isotones were performed using the relation \cite{BM71}
\begin{equation}\label{four_point}
	\Delta_{nn}^{(4)}(N)=\frac{(-1)^N}{2}[-S_n(N+1)+2S_n(N)-S_n(N-1)],
\end{equation} 
which serves as an averaging of $\Delta_{nn}(N, Z)$ for two neighbouring nuclei. All the data were taken from AME2020 \cite{AME2020}.

Fig.~\ref{exp} shows the experimental spectra in even neutron-rich nickel isotopes and isotones $N=50$ for comparison. We should point out the distinct spectra of $^{90}$Zr and $^{68}$Ni as opposed to other nuclides in the chains: the first excited states $2^+$ and $4^+$ there lie higher than the corresponding states in adjacent nuclei, although this effect is not as prominent as in other magic nuclei.  The spectra of these and the rest of the isotopes demonstrate the trademark of GSM coming from pairing of like nucleons on $j=9/2$: even-spin states with a small splitting around $~100$~keV between $8^+$ and $6^+$. The values of $\Delta_{\tau\tau}^{(4)}$ ($\tau=n, p$) correspond to the position of state $8^+$ and are around 0-250 keV above it, and not just for $^{70}$Ni and $^{92}$Mo with two nucleons above the closed shell, but for isotopes with more nucleons as well.

This is in agreement with seniority scheme, where the total splitting of the GSM part with $\nu=2$ (where $\nu$ is seniority number, the number of unpaired nucleons) remains constant from isotope to isotope throughout the filling of the whole subshell. The possible values of the total angular momentum $J$ for nuclei with $N$ valence nucleons on the outer $j=9/2$ subshell, among which $\nu$ are unpaired, are given in table~\ref{tab9_2}. As $j$ grows, the sets of possible values of $J$ for $\nu>2$ become more complex, one of the first such calculations for $j>7/2$ performed in \cite{GM55}.

\begin{center}	
	\tabcaption{Summation of angular momenta $j = \frac{9}{2}$}
			\begin{tabular*}{80mm}{cc}
			\hline
			\,\,\,\,\,\,$v$\,\,\,\, &	Total angular momentum $J$ \\ 
			\hline
			$0$ & $0$ \\
			$1$ & $\frac92$ \\
			$2$ & $2,4,6,8$ \\
			$3$ &\,\, $\frac32,\frac52,\frac72,\frac92,\frac{11}{2},\frac{13}{2},\frac{15}{2},\frac{17}{2},\frac{21}{2}$ \,\,\\
			$4$ & $0,2,3,4^2,5,6^2,7,8,9,10,12$ \\
			$5$ &\,\,\,\,\,\, $\frac12,\frac52,\frac72,\frac92,\frac{11}{2},\frac{13}{2},\frac{15}{2},\frac{17}{2},\frac{19}{2},\frac{25}{2}$ \,\,\\
			[4pt]
			\hline
		\end{tabular*}\label{tab9_2}
	\end{center}

In seniority scheme, the states with higher seniority $\nu$ are also degenerate. The splitting occurs in calculations based on the non-degenerate set of $\nu=2$ levels, which can either be found using different models or taken from experiment. The wave functions and energies of a three-nucleon system can be expressed as linear combinations and wave functions or energies of the system with two nucleons with an added third nucleon. For energies of the excited states of the three-nucleon system, the following formula holds:
\begin{align}
	E(J_3)=\sum_{J_2}[(jj)J_2 j J_3 |\} j^3 J_3]^2 E(J_2).
\end{align}
Here, $[(jj)J_2 j J_3 |\} j^3 J_3]$ are the fractional parentage coefficients (FPC). The calculated spectrum of the system of three nucleons  can then be similarly used to find the energies of the system with more nucleons on the given subshell by means of the iterative procedure. In our calculations we used the FPC values from \cite{Bayman}.

The seniority scheme was first proposed by Racah and Flowers \cite{R42,R43,F52} for the cases $j \le 7/2$. However, seniority number may not be necessarily conserved for higher values of $j$. When it comes to the case of $j=9/2$, there appear pairs of $J=4$ and $J=6$ states with $\nu=4$ which may partly be mixed with corresponding $\nu=2$ states. At the same time, there remain two distinct $J=4$ and $J=6$ states, for which seniority remains a good quantum number no matter the form of nucleon-nucleon interaction \cite{EZ06, Z07, IH08, IH14}. Analytical and numerical studies in this regard showed that these states are indeed reproduced within Bayman's FPC scheme \cite{Bayman} with accuracy of 99\% and above \cite{Qi11}.

\section{Results}

Seniority model together with the $\delta$-force approximation for pairing was used describe the GSM spectra in $^{70-76}$Ni, with $^{68}$Ni taken as the doubly magic nucleus ($Z=28$). Nuclei in this range of masses and neutron excess are commonly treated using the shell model configuration mixing taken into account. For isotopes studied in this work, such calculations within $pfg_{9/2}d_{5/2}$ model space showed that the occupation of $\nu g_{9/2}$ in $^{68}$Ni is affected on a rather negligible scale of around 4.7\% compared to isotones $N=40$ \cite{CC14}. Experimental data for $^{69-77}$Ni also show that it is the $\nu g_{9/2}$ single particle state that is primarily filled along this chain of isotopes. This is evident from the ground states of odd isotopes having the spin of $J^\pi=9/2^+$, as well as from the observed ground state multiplets coming from pairing of neutrons on  $j=9/2$ subshell. For this reason, we treated the isotopes $^{70-76}$Ni within the model space including just the subshell $g_{9/2}$. Similar calculations were performed in isotones$N=50$ with valence protons in $g_{9/2}$.

\subsection{Two nucleons on $j=9/2$}

Fig.~\ref{nu2}a presents the comparison between experimental and calculated spectra in isotopes $^{70}$Ni ($^{76}$Ni) with 2 nucleons or (2 holes) in $1g_{9/2}$. Notably, higher momentum states $6^+$ and $8^+$ are reproduced most accurately within the approximation of $\delta$-potential, the energies of $4^+$ and $2^+$ are overestimated by 250-300 keV and $\sim$ 1 MeV respectively. A similar situation is observed in spectra of $^{92}$Mo and $^{96}$Cd (Fig.~\ref{nu2}b).

\begin{center}
\includegraphics[width=85mm]{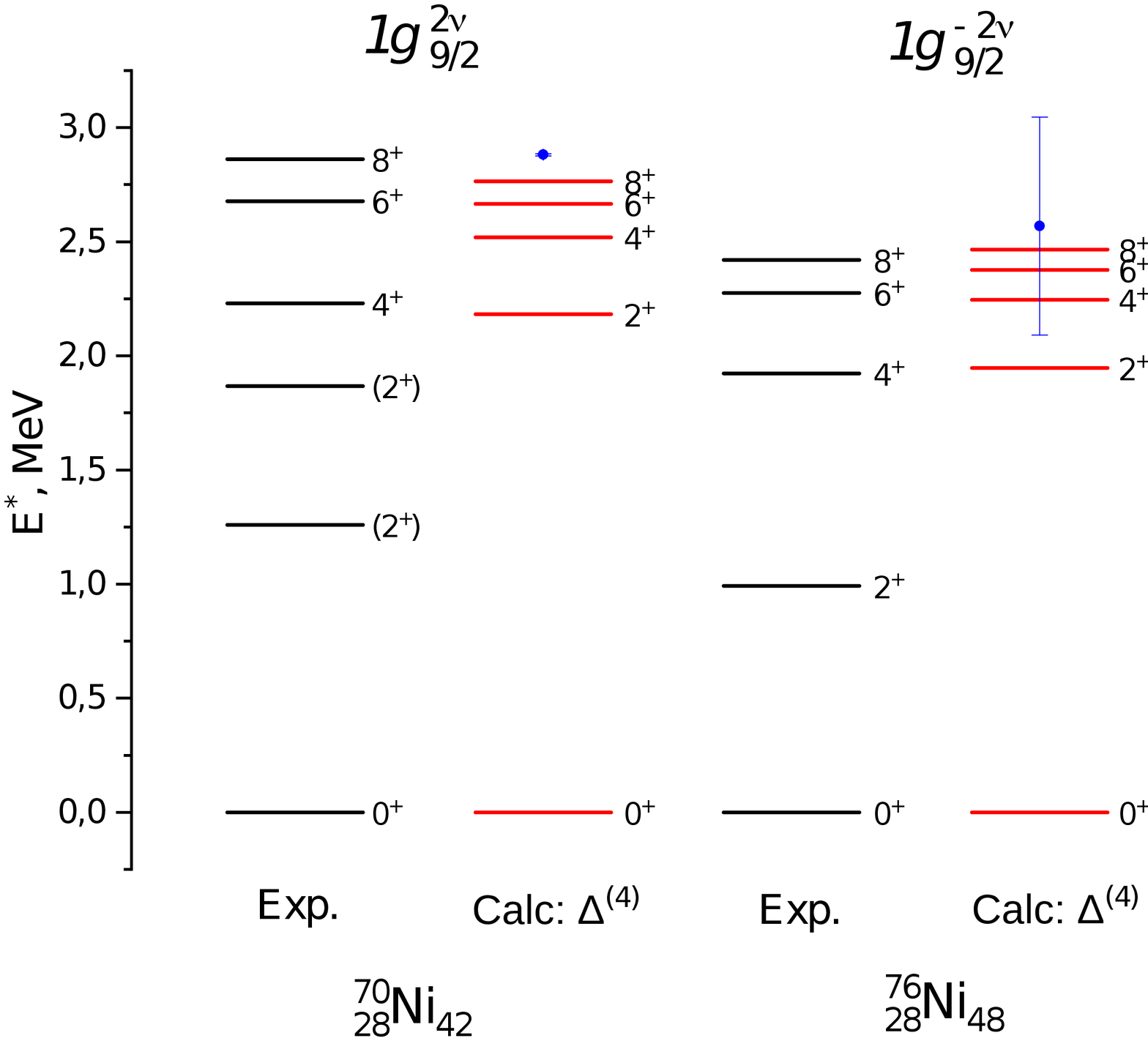}   
\vspace{30pt}
\includegraphics[width=83mm]{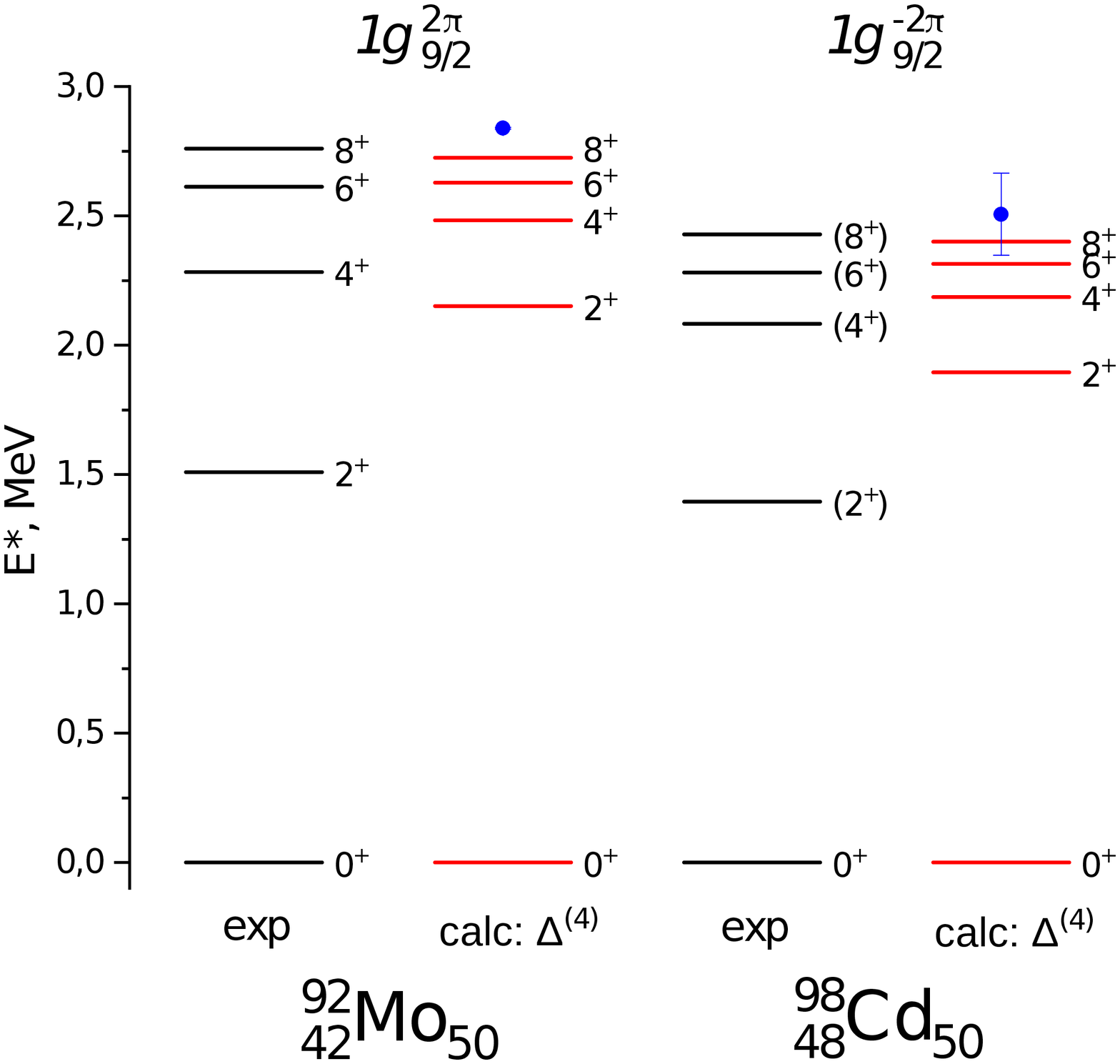} 
\figcaption{Experimental (left) and calculated (right) low-energy level schemes of nickel isotopes  $^{70,76}$Ni (top) and $N=50$ isotones $^{92}$Mo, $^{98}$Cd (bottom). The dots represent the value of $\Delta_{nn}^{(4)}$ in nickel isotopes or  $\Delta_{pp}^{(4)}$ in $N=50$ isotones . 
	Experimental data are from \cite{GM16} ($^{70}$Ni), \cite{FS07} ($^{76}$Ni), \cite{B12} ($^{92}$Mo) and \cite{C20} ($^{98}$Cd). See text for details.}
\label{nu2}
\end{center}

It should be noted that in decomposition of residual interaction by Legendre polynomials
$$v(r_{12})=\sum_l f_l(r_1,r_2)P_l(\cos \Theta_{12}),$$ 
the values of coefficients $f_l$ for $\delta$-interaction are of the form:
$$f_l = \delta(r_1-r_2)\frac{2l+1}{4\pi r_1^2},$$
meaning larger values of $f_l$ correspond to larger $l$.

Terms with lower $l$ are responsible for long-ranged interaction and collective nuclear effects \cite{Lane}. It is for this reason $\delta$-interaction works best when describing states with higher angular momentum. On contrary, for states $J^\pi=2^+$, the first such state in experimental spectra often lies at energies of around the BCS gap $\Delta_\tau = \Delta_{\tau \tau}/2$, whereas in $\delta$-approximation $E^*(2^+)/\Delta_{\tau \tau}^{(4)}=0.75$. In the region of medium and light nuclei these states are often reproduced via configuration mixing.

It should be noted that the first $2^+$ states lie below the value of $\Delta_{\tau\tau}/2$ in $^{70}$Ni and $^{76}$Ni. Specifically, for $E^*(2^+)/\Delta_{nn}^{(4)} = 0.44$ for $^{70}$Ni and 0.39 for $^{76}$Ni. For protons, the corresponding relation $E^*(2^+)/\Delta_{pp}^{(4)}$ is higher: 0.53 for $^{92}$Mo and 0.56 for $^{98}$Cd. Later we will see this resulting in significant differences in the structure of spectra with higher seniority.

\subsection{Four nucleons on $j=9/2$, seniority $\nu=2,4$}

In accordance with seniority model, the spectrum of $^{72}$Ni with four neutrons on $1g_{9/2}$ should include a set of $\nu=2$ states (observed in $^{70}$Ni) complemented by levels with $\nu=4$. Considering the location of $\nu=4$ states depends on the energies of $\nu=2$ levels, it is important to verify the sensitivity of the former, in particular, to the energy of the $J=2^+ (\nu=2)$ overestimated within $\delta$-potential approximation. The corresponding calculation is shown on Fig.~\ref{7274Ni} as the first spectrum. All the $\nu=4$ here and on are shown with coloured lines. Notably, the energies of the second $J^\pi=4_2^+, 6_2^+$ states are overvalued by $\sim$1 MeV. We assumed that this overestimation primarily comes as a result of inaccuracy in the energy of $J^\pi=2^+ (\nu=2)$ and performed a calculation with the fixed energy value of this state taken from experiment. And indeed, fixing this state results in the increase in the splitting of the $\nu=4$ spectrum. Furthermore, $^{72}$Ni observes the shift of $J^\pi=4^+,6^+ (\nu=4)$ levels below the corresponding $\nu=2$ states. In other words, an inversion of excited states occurs as per seniority model. Such an anomalous order of levels was described in the article \cite{MBW18}, the authors of which drew conclusions about the structure of the spectra based on intensity of transitions between different levels.

\end{multicols}
\begin{center}
\includegraphics[width=110mm]{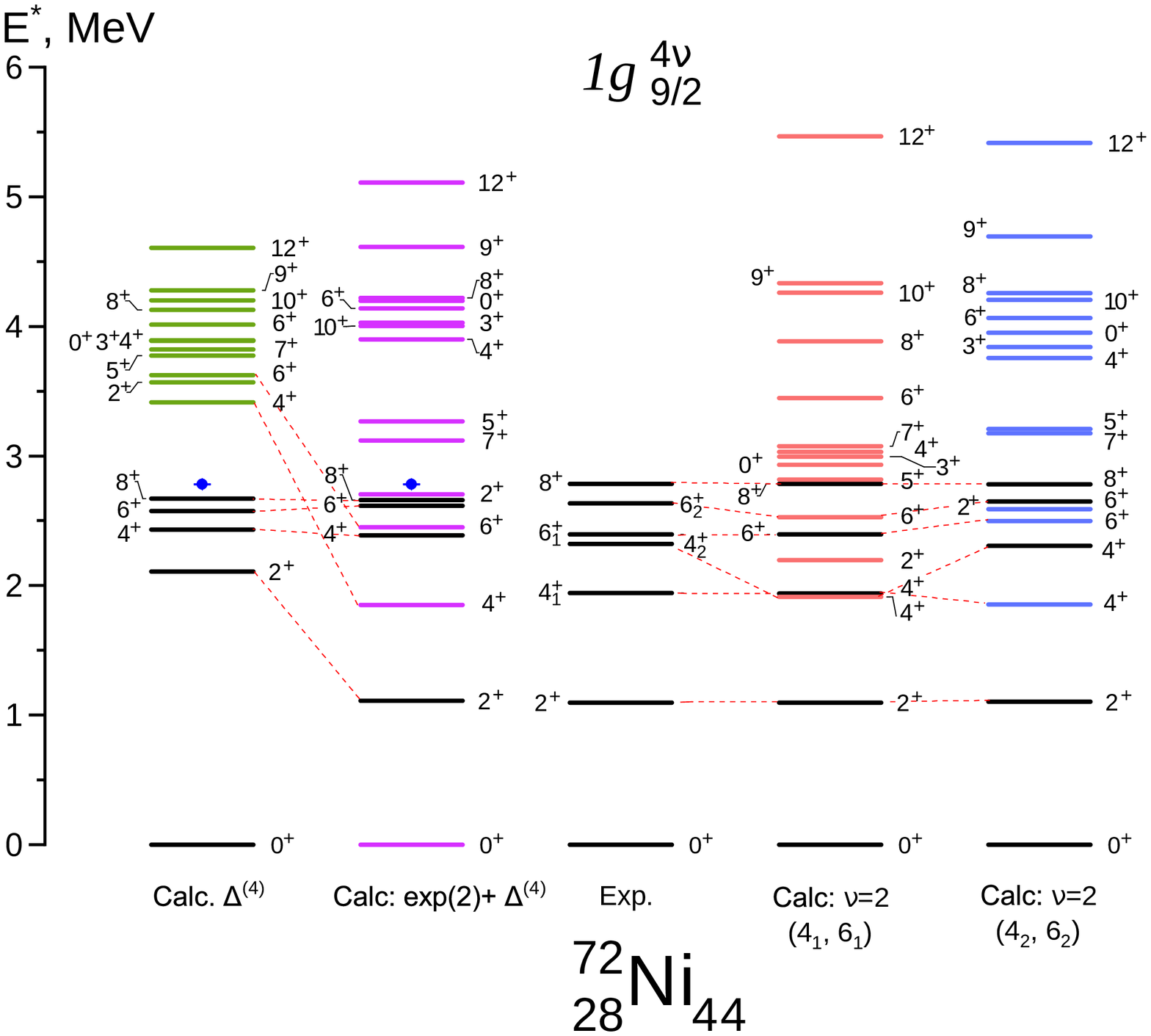}   
\includegraphics[width=110mm]{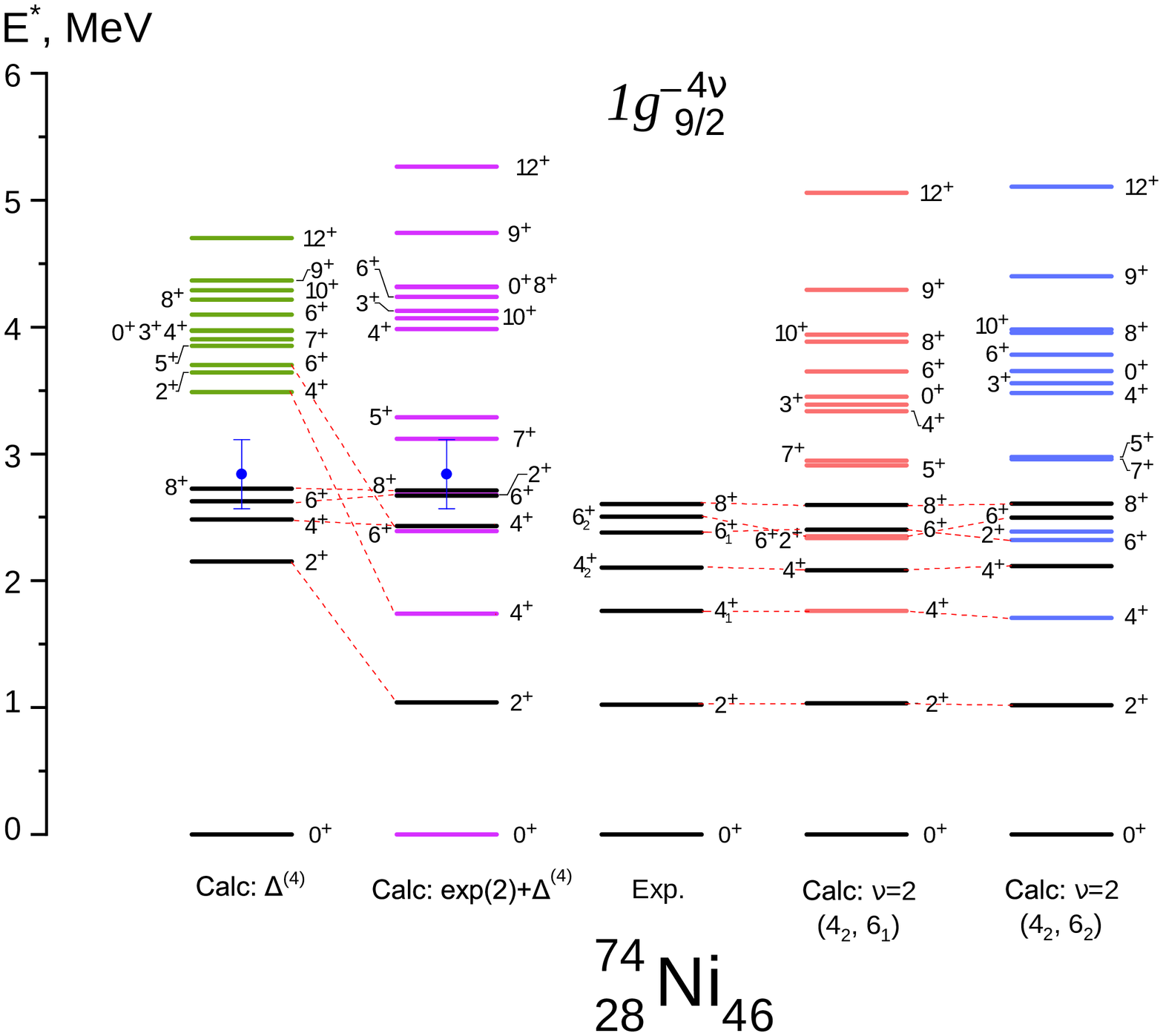} 
\figcaption{Low-energy level schemes of nickel isotopes  $^{72}$Ni (top) and $^{74}$Ni (bottom). States with senority $\nu=2$ are shown in black, states $\nu=4$ --- by colored lines. The experimental spectrum is shown in the center (data from \cite{AS06, AS08}).
Left: first column --- calculations in the $\delta$-approximation based on $\Delta_{nn}^{(4)}$ (marked with a dot), the second one --- the same using the experimental value $2^+$. On the right are calculations using different experimental values $4^+$ and $6^+$ as states with senority $\nu=2$. }
\label{7274Ni} 
\end{center}
\begin{multicols}{2}

To make more objective predictions regarding which experimental states ($4_1$ or $4_2$, $6_1$ or $6_2$) correspond to seniority $\nu=2$ and $\nu=4$, we carried out several calculations with all the $\nu=2$ state energies fixed and taken from experimental data, for each of the combinations of states $4_i$, $6_j$. Examples of such calculations are also shown on Fig.~\ref{7274Ni} as the fourth and fifth spectra. Namely, the fourth spectrum was obtained in the assumption experimental states $4_1^+, 6_1^+$ are characterized by seniority $\nu=2$ and their energy can thus be used to reproduce the spectrum $\nu=4$; similarly, $4_2^+, 6_2^+$ levels were taken as $\nu=2$ on the fifth spectrum. Comparison of these spectra to experiment shows that $4_2^+$ should likely be the $\nu=2$ state as discussed earlier. Data on decay of $^{72}$Ni supports this claim. At the same time, it remains uncertain which of the $6_{1,2}$ belongs to $\nu=2$ part of the spectrum, as both calculations reproduce these levels.

Similar calculations were carried out for $^{74}$Ni with four holes on $1g_{9/2}$. We drew similar conclusions for the $4^+$ levels in this nucleus: the lower of the two is the $\nu=4$ state. The $6^+$ states, on the other hand, can barely be resolved seniority-wise due to even smaller spacing between them. And just like in $^{72}$Ni, the location of higher momentum levels also has a strong dependence on $E(2^+, \nu=2)$: in both of these nuclides the energy of $12^+$ is estimated at 5 -- 6 MeV. 

\end{multicols}
\begin{center}
\includegraphics[width=95mm]{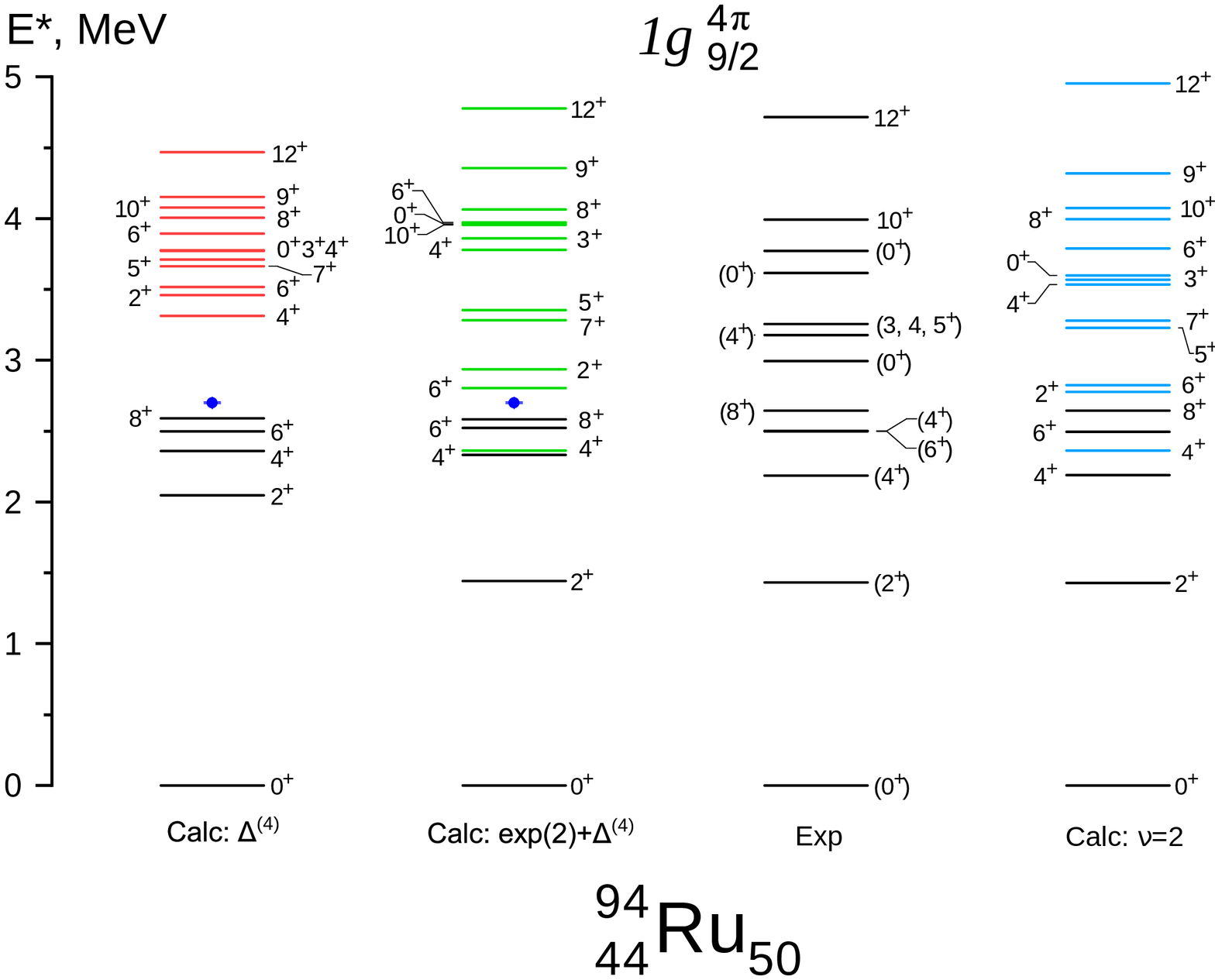}   
\includegraphics[width=95mm]{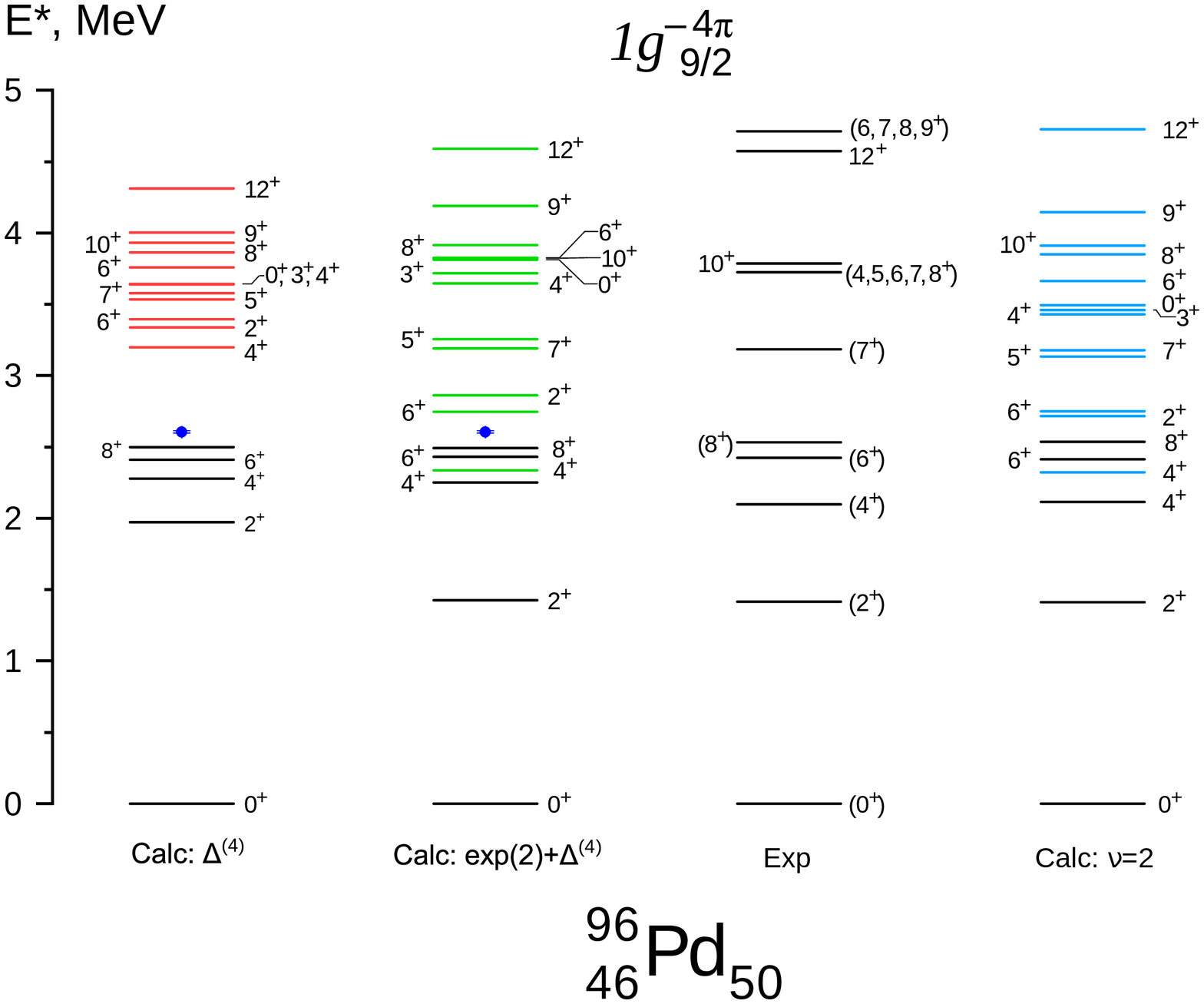} 
\figcaption{Low-energy level schemes of $N=50$ isotones  $^{94}$Ru (top) and $^{96}$Pd (bottom). See caption of Fig.~\ref{7274Ni}.  Experimental data taken from \cite{M16, MBW18}.}
\label{94Ru96Pd}
\end{center}
\begin{multicols}{2}
The absence of data on the position of the levels with $J^{max} = 12^+$ in the spectra of $^{72}$Ni and $^{74}$Ni does not allow us to compare the total splitting of the GSM with the experiment.  From this point of view, it is all the more important to compare the corresponding calculations for isotones $N=50$. Low energy spectra of semi-magic isotopes $^{94}$Ru and $^{96}$Pd exhibits the example of the dominance of isovector paring in accordance with the seniority scheme \cite{RR01,CG11}. 

Experimental spectra of $^{94}$Ru and $^{96}$Pd are compared to our calculations on Fig.~\ref{94Ru96Pd}. The $\nu=4$ spectra for protons in the $N=50$ isotone chain show an important difference: there are still two $4^+$ levels within the energy boundaries of the $\nu=2$ part of the GSM (energies up to $8_1^+$ state), yet only one $6^+$ level. Unlike nickel isotopes, in this case the first $4^+_1$ state belongs to senority $\nu=2$, which turns out to be important in interpreting the observed E2 transitions properties in these isotopes \cite{Qi17}. Calculations $\delta$-potential approximation showed that fixing just the  energy of $2^+$ is enough to reproduce the general order of excited states known experimentally and achieve a significantly better correspondence between the energies of particular levels, especially when it comes to high momentum states $12^+$ and $10^+$.

The correspondence between the $0^+ - 2^+$ splitting and location of $(6^+, \nu=4)$ was discussed earlier, for example, in \cite{L04}. The authors of the article drew a connection with the differences between the cores $^{90}$Zr and $^{68}$Ni. The model assumes the presence of a closed core and is thus rather simplified in the given case, as neither $^{68}$Ni nor in $^{90}$Zr possesses a distinctly magic number of neutrons or protons (40 in this case). For example, in microscopic calculations within dispersive optical potential, the estimate of $\nu g_{9/2}$ occupation in $^{68}$Ni is at 20\% \cite{B11Ni}, the occupation of $\pi g_{9/2}$ in $^{90}$Zr: 14\% \cite{B15Zr}. When describing Ni isotopes within shell model and comparing then to $N=50$ isotones, the authors of \cite{L04} point out the differences between the order of single particle states in the $fp$-shell and the differences between the spectra coming as a result. None the less, our analysis shows that GSM spectra of nuclei with like nucleons on $g_{9/2}$ may present an excellent opportunity to study the conservation of seniority for nucleons in $j=9/2$ state.

\subsection{Three nucleons on $j=9/2$, $\nu = 3$}

Experimental spectra of odd nickel isotopes under consideration also aren't as well studied as those of more stable nuclides. The spectrum of $^{71}$Ni, for example, includes 5 states in the range of 1.3~MeV, whereas 8 $\nu=3$ states should be observable within the seniority scheme.

\end{multicols}
\begin{center}
\includegraphics[width=120mm]{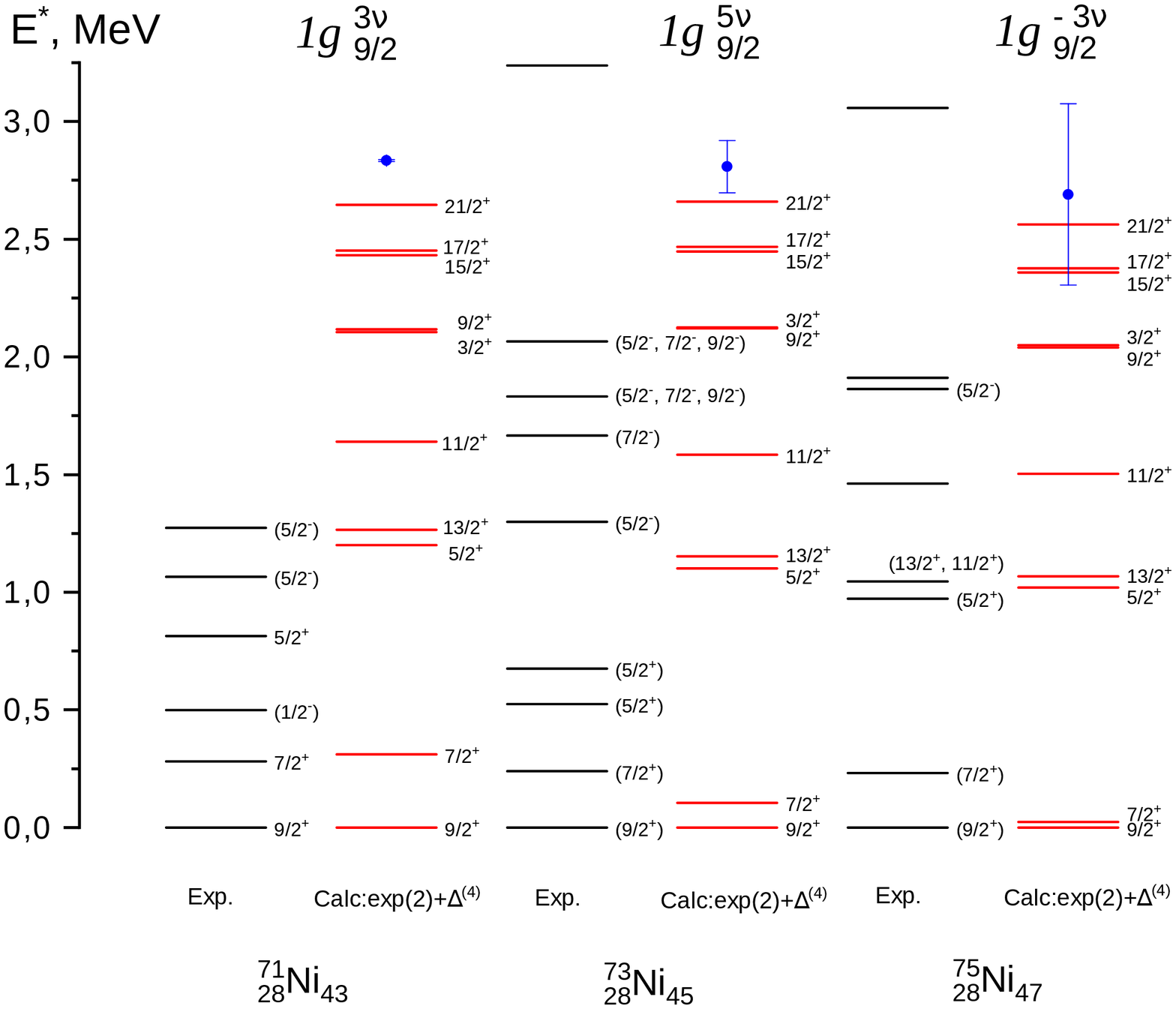}   
\figcaption{Experimental (left) and calculated (right) low-energy level schemes of nickel isotopes $^{71,73,75}$Ni. Experimental data taken from \cite{ENDF_Ni71} ($^{71}$Ni), \cite{GGM20} ($^{73}$Ni), \cite{EMN21} ($^{75}$Ni)). The calculated states with $\nu=3$ were obtained in the $\delta$-force approximation using the experimental energy value  of the $2^+_1$ state in the neighboring even-even isotope in the spectrum of senority $\nu=2$.}
\label{nu3}
\end{center}
\begin{multicols}{2}

The splitting of this GSM is equal to $4/5\Delta_{nn}$ which equals roughly 2.27~MeV for $^{71}$Ni, and the maximal spin is $J=\frac{21}{2}$. Some impressive experimental data on $^{73}$Ni and $^{75}$Ni was recently obtaned on RIBF (RIKEN) \cite{GAD20,EMN21} and NSCL (MSU) \cite{GGM20}, including new information on the state $\frac{13}{2}^+$.

Earlier we showed that the most accurate estimates on energies of $\nu=4$ levels in even nuclei are attained through the combined use of $\delta$-interaction approximation and the fixed experimental energy of the $2^+$ state. Similar calculations can be performed in the case of odd isotopes if the energy value of $J=2^+$ state is taken from the neighbouring even isotope. The corresponding calculated spectra are presented on Fig.~\ref{nu3}.

It seems most reasonable to base the calculations of $\nu=3$ levels in $^{71}$Ni on the energy of level $2^+$ in $^{70}$Ni as nucleus with 2 nucleons on the shell. In reality, there are 2 such levels in the experimental spectrum of $^{70}$Ni which results in ambiguity as to which belongs to the GSM coming from pairing. Two separate calculations were carried out for this reason, one based on the energy of either state $2_1^+$ or $2_2^+$. We found that the only known experimental states with positive parity in $^{71}$Ni, $J=7/2^+, 5/2^+$, are best reproduced when the energy of $J=2_1$ is used, proving it's $\nu=2$ nature and affiliation with the GSM.

The $1g_{9/2}$ subshell of $^{73}$Ni is half filled, and strictly speaking, its spectrum should be made up of $\nu=1,3,5$ levels. We chose to limit our calculations to seniority $\nu=1,3$ and use the experimental energy of the $J=2$ level in $^{72}$Ni. For $^{75}$Ni with three holes, similar calculations were based $E(2^+)$ in $^{76}$Ni. Here, reasonable agreement was achieved for the states $\frac{5}{2}^+, \frac{13}{2}^+$.

Analogous studies of the $^{93}$Tc, $^{95}$Rh and $^{97}$Ag
spectra may also prove fruitful, as more experimental data on higher spin states are available for these nuclei (see Fig.~\ref{nu3_50}). In comparison with our previous calculations \cite{Step}, the use of the experimental value energy of the state $J^P=2^+ (\nu=2)$ from the spectrum of the neighboring even-even isotope made it possible to better reproduce both the position of $\frac{7}{2}^+$ state and levels with $J$ high value.  It is worth noting the reliable reproduction of the overall splitting aka the energy of the $J^\pi=21/2^+$ state.

\end{multicols}
\begin{center}
\includegraphics[width=120mm]{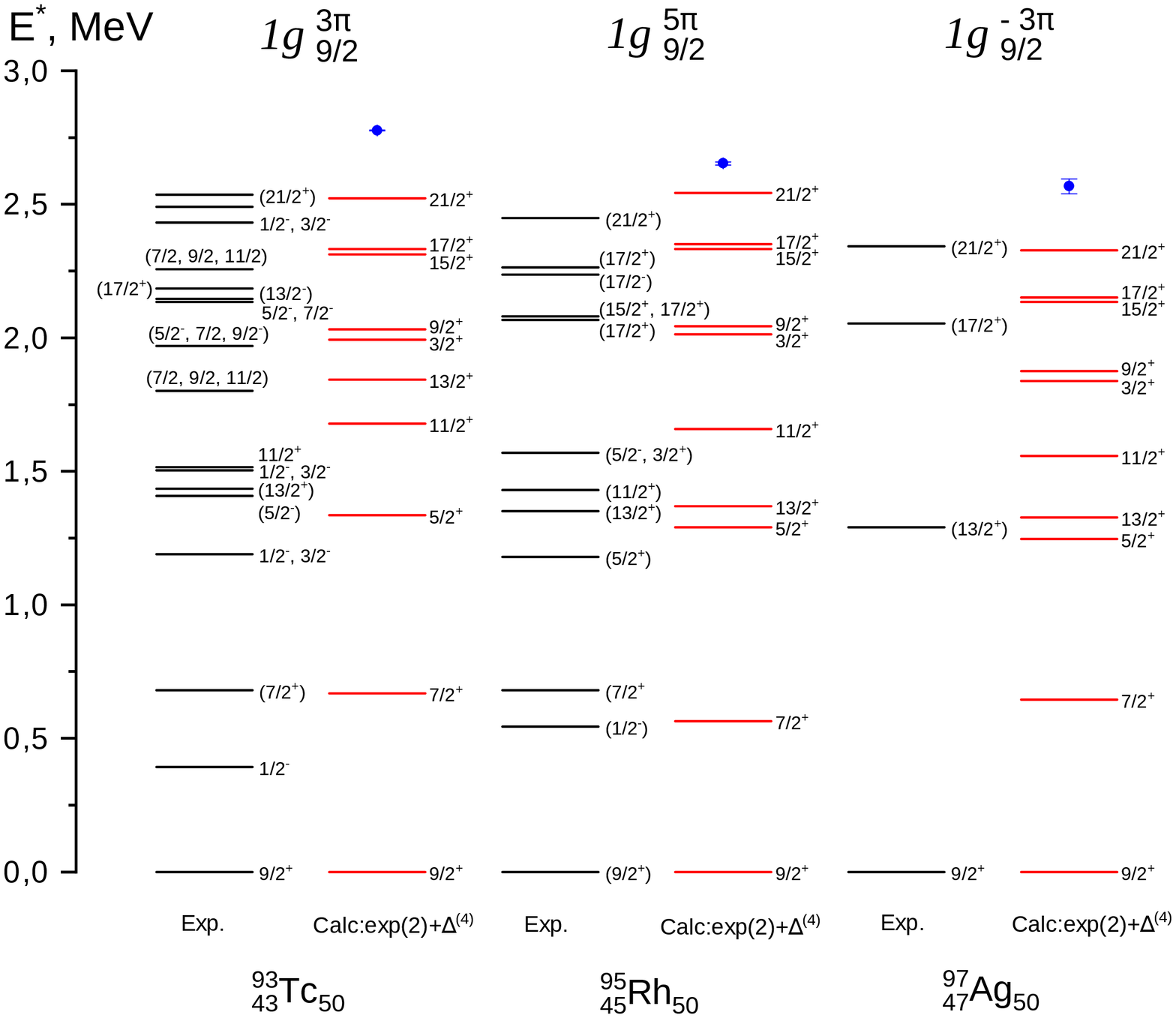} 
\figcaption{Experimental (left) and calculated (right) low-energy level schemes of $N=50$ isotones $^{93}$Tc, $^{95}$Rh and $^{97}$Ag. Experimental data taken from \cite{B11} ($^{93}$Tc), \cite{BMS10} ($^{95}$Rh), \cite{N10} ($^{97}$Ag). See caption of Fig.~\ref{nu3}.}
\label{nu3_50}
\end{center}
\begin{multicols}{2}

As seen from our estimates for $\nu=3,4$ levels in various isotopes, GSM splitting can be very sensitive to the location of $(J^\pi=2^+, \nu=2)$. The structure of the whole $\nu=0,2,4$ spectrum as a function of $E(2^+)/\Delta_{\tau \tau}$ is shown on Fig.~\ref{v=4}. As seen from the graph, the relative energy of $2^+$ has the strongest influence on $4^+$ and $6^+$ in $\nu = 4$ spectrum. The second $4^+, 6^+$ has a negligibly small dependence on $\nu=2$ states and does not change its position, as neither does $8^+$, which speaks of seniority being a good quantum number for these levels. Among the odd-spin states, $3^+$ is the least sensitive to the energy of $(J=2,\nu=2)$.

\begin{center}
	\includegraphics[width=8cm]{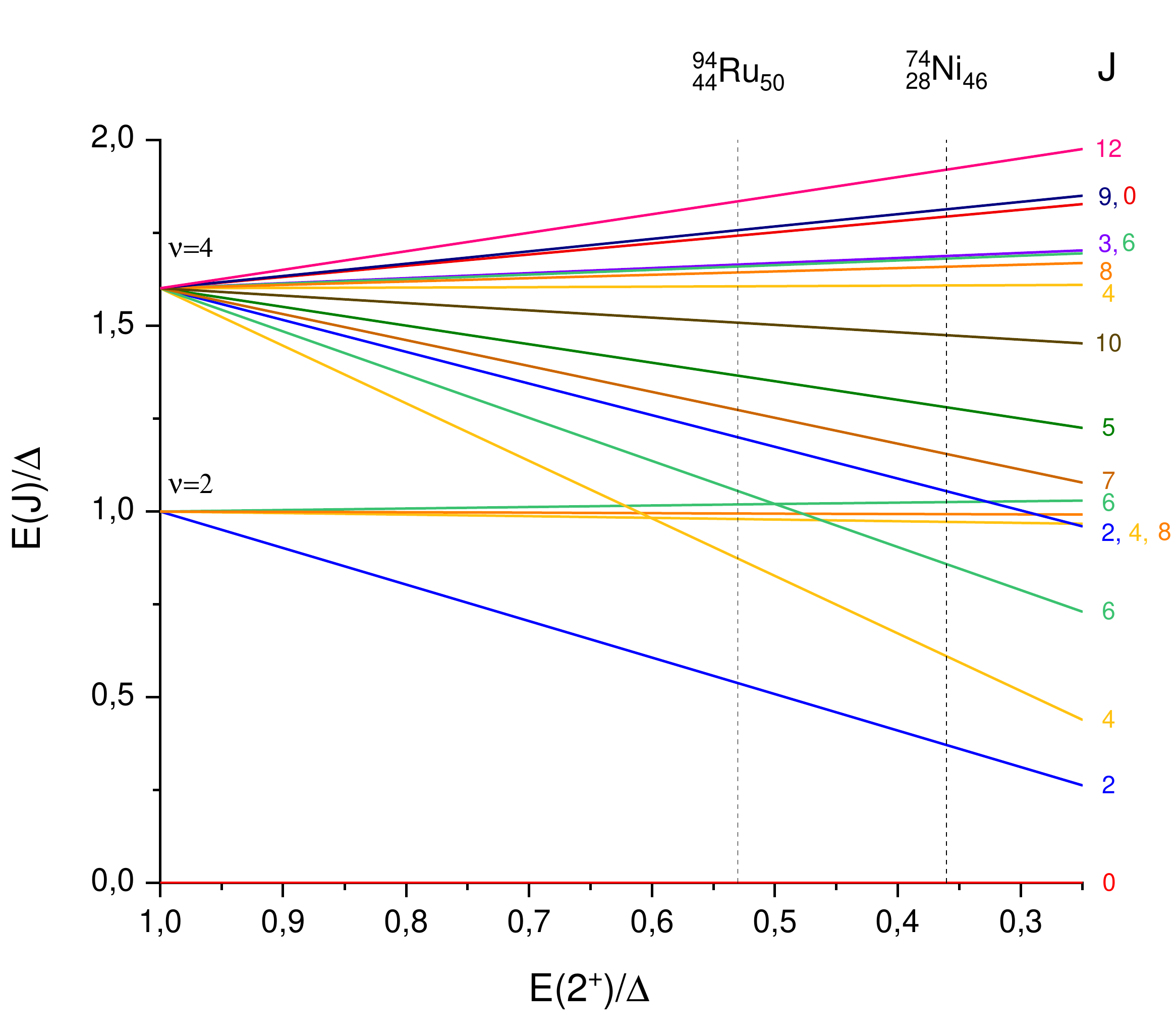}
	\figcaption{Seniority $\nu=4$ spectra versus energy of $2^+$ in $\nu = 2$ spectrum.}
	\label{v=4}
\end{center}

The value of $E(2^+)/\Delta_{\tau \tau}$ also has a significant effect on the overall GSM splitting and the difference between the energies of high momentum levels $12^+$ and $10^+$. As for the structure of the multiplet, we shall recall that the $\delta$-force approximation yields the value of $E^*(2^+)/\Delta_{\tau\tau}\approx 0.75$ for the nucleons on $j=9/2$ shell --- the value at which, in accordance with Fig.~\ref{v=4}, no significant changes in the GSM structure happen. For both isotones $^{94}$Ru and $^{96}$Pd this relation is equal to around 0.53, which, as seen from the graph, results in $(4^+$ $\nu=4)$ 'dropping' below the pairing energy in the region of $\nu=2$ part of GSM. For nickel isotopes this relation is even lower (at 0.39 MeV for $^{72}$Ni and 0.36 for $^{74}$Ni) - and indeed, the second $6^+$ state also shifts down to $\nu=2$ energies.

\section{Conclusions}

Pairing correlations have a significant influence on the properties of atomic nuclei. The multiplet of excited states as one of its consequences depends on many factors including the form of residual interaction, shell structure of the nucleus and its deformation. Nevertheless, even a simplified model that does not take configuration mixing and long-ranged part of nucleon-nucleon interaction into account, may give us a good understanding regarding the structure of low-lying excited states in a nucleus with an isolated subshell.

The approximation of $\delta$-interaction for nucleon pairing was employed to treat the $\nu=2$ part of GSM spectra in $^{70-76}$Ni and isotones $N=50$ with valence nucleons on $g_{9/2}$ subshell. Mass relations were shown to give a reasonable estimate for GSM splitting; ($J^\pi=6^+,8^+$, $\nu=2$) states in even nuclei were reproduced with an error of no more than 0.1 MeV and ($J^\pi=4^+$, $\nu=2$) were reproduced with an error of no more than 0.5~MeV. This approximation gives a significantly higher energy value $E(2^+)$ due to presence of configuration mixing and a long-range part of nucleon interaction.

A correct account of $(J^\pi=2^+$, $\nu=2)$ (whether through use of more realistic interaction or just direct fixing of experimental values) yields some interesting effects. The remarkable dependence of $\nu = 4$ GSM splitting on location of this state results in a shift of $12^+$ in isotopes with 4 nucleons (or 4 holes) on the $j=9/2$ subshell to higher energies by 300-500 keV, which allows for a better description of this and other high momentum states, particularly, in $^{94}$Ru and $^{96}$Pd. Two states $(J=4,6,\nu=4)$, on the contrary, drop down, and the shift is more significant the lower $(J=2, \nu=2)$ is. This effect is especially striking in $^{72,74}$Ni, where the very low (in relation to GSM splitting) $2_1$ level brings about the inversion of the order of $4^+$ states with different seniority. The presence of such an inversion is not so obvious for $6^+$ states, and it's highly likely seniority is not a good quantum number for these levels.

\end{multicols}

\vspace{10mm}

\vspace{-1mm}
\centerline{\rule{80mm}{0.1pt}}
\vspace{2mm}

\begin{multicols}{2}

\end{multicols}

\clearpage

\end{CJK*}
\end{document}